\documentclass[a4paper,14pt]{article} 

\usepackage[T2A]{fontenc}
\usepackage[cp1251]{inputenc}
 \usepackage[russian]{babel}
\usepackage{amssymb,amsfonts,amsmath,mathtext,cite,enumerate,float} 
\usepackage{graphicx}

 \makeatother

 \usepackage{geometry} 
\begin{document}
\begin{center}
\Large \textbf{ Решение задачи о релаксации для газа с функцией распределения, зависищей от модуля скорости} \\*
\vspace{0.5cm}
\large \text{George Arabuli} \\*
\vspace{0.5cm}
\large \textit{Moscow Institute of Physics and Technology (State University)}
\end{center}

\section*{Abstract} В работе предложен метод решения кинетического уравнения Больцмана, оcнованный на построении его дискретной консервативной модели. Дискретный аналог интеграла столкновений представляется в виде свертки тензора столкновений, не зависящего от функции начального распределения, с тензором с компонентами из средних плотностей в ячейках. Численная реализация дискретной модели продемонстрирована на задачи об изотропной релаксации газа для модели твердых сфер.
\newline
Главной особенностью   метода является независимость компонент тензора столкновений от функции распределения. Таким образом, компоненты тензора столкновений вычисляются один раз для разных начальных функций распределения, что значительно увеличивает быстродействие предложенного метода.
\section{Введение}
Движение разреженного газа описывается функцией распределения $f(t,\vec{x},\vec{u})$, удовлетворяющей кинетическому уравнению Больцмана  \cite{baran}:
\begin{multline}
\label{inbaran}
\frac{\partial{f}}{\partial{t}} + \sum\limits_{i = 1}^{3}u_i \frac{\partial{f}}{\partial{x_i}} = \int\int f(\vec{u_1})f(\vec{u_2}) T(\vec{u_1},\vec{u_2},\vec{u}) \cdot  \\
\cdot \sigma(|\vec{u_1}-\vec{u_2}|) |\vec{u_1}-\vec{u_2}|d\vec{u_1}d\vec{u_2} - f(\vec{u}) \int f(\vec{u_1}) \sigma(|\vec{u}-\vec{u_1}|) |\vec{u}-\vec{u_1}|d\vec{u_1}
\end{multline}

Здесь $x_i$, $u_i$ - декартовы координаты физического и скоростного пространства, $t$ - время,\newline 
$u_1$, $u_2$ - скорости сталкивающихся частиц, \newline 
$\sigma(|\vec{u_1}-\vec{u_2}|)$ - полное сечение взаимодействия, \newline
$T(\vec{u_1},\vec{u_2},\vec{u}) $ - ударная трансформанта. 
\newline
Для модели твердых сфер диаметра $d$ имеем:
$\sigma(|\vec{u_1}-\vec{u_2}|) = \pi d^2$ и 
\begin{eqnarray}
\title{tsfers}
T(\vec{u_1},\vec{u_2},\vec{u}) = \frac{1}{\pi |\vec{u_1}-\vec{u_2}|}\delta (|\vec{u}-\frac{\vec{u_1}+\vec{u_2}}{2}| -\frac{|\vec{u_1}-\vec{u_2}|}{2})
\end{eqnarray}
Для построения консервативного решения этого уравнения можно разбить область на $N$ Элементарных ячеек объемом $\Delta V = \Delta u_1\Delta u_2\Delta u_3$, пронумировах их номерами $k$ от $1$ до $N$. Тогда функцию распределения $f(t,\vec{x},\vec{u})$ можно представить в виде:
\begin{eqnarray}
\label{zambar}
f(t,\vec{x},\vec{u}) = \sum\limits_{k = 1}^{N} f_k \delta(|\vec{u}-\vec{u_k}|)
\end{eqnarray}
$f_k = f(t,\vec{x},\vec{u_k})$ - значение функции распределения в k-ой ячейки. Подставляя выражение (\ref{zambar}) в интеграл (\ref{inbaran}),  и переходя от уравнения на функцию распределения к уравнению на число частиц $n_k = f_k\cdot\Delta V$, получаем систему уравнений:
\begin{eqnarray}
\label{finbaran}
\frac{\partial{n_k}}{\partial{t}} + \sum\limits_{i = 1}^{3}u_{ki} \frac{\partial{n_k}}{\partial{x_i}} = \sum \limits_{i=1}^N\sum \limits_{j=1}^N A_{ij}^{k}n_in_j - n_k\sum \limits_{j=1}^N B_j^kn_j
\end{eqnarray}
Коэффициенты $A_{ij}^{k}$ и $ B_j^k$ не зависят от функции распределения, поэтому их достаточно вычислить один раз. Вычисление коэффициентов в трехмерном случае представляет значительные вычислительные трудности. Поэтому, вычисление этих коэффициентов разумнее производить для более простых случаев: функция распределения зависящая только от модуля скорости или от продольной скорости и модуля поперечной.
\newline
 Зная коэффициенты $A_{ij}^{k}$ и $ B_j^k$ мы сможем найти значения $n_k$  на следующем временном слое. Однако не будет выполнен закон сохранения числа частиц и закон сохранения энергии. Для обеспечения выполнения законов сохранения мы будем вынуждены корректировать полученные значения $n_k$.
\newline
В работе \cite{aristov} автором предложен метод построения консервативного решения дискретного уравнения Больцмана с коррекцией чисел заполнения $n_k$ на каждом шаге по времени. 
\newline
В следующих разделах будет предложен метод построения консервативного решения дискретного уравнения Больцмана для функции распределения, зависящей только от модуля скорости. Особенностью этого метода будет являться однократная корректировка коэффициентов $A_{ij}^{k}$ и $ B_j^k$ и отсутствие корректировки чисел заполнения $n_k$  .
\section{Постановка задачи}
\subsection {Уравнение Больцмана}
Исходное уравнение \cite{rykov}:
\begin{multline}
\label{initial}
\frac{\partial{f(t,V)}}{\partial{t}} = 
2\pi^2 d^2 (
 \int\limits_0^V \int\limits_{\sqrt{V^2-x^2}}^V f(x)f(y)\frac{4\sqrt{x^2+y^2-V^2}}{V}ydydx +\\+
 \int\limits_V^{\infty}f(x)2xdx \cdot \int\limits_V^{\infty}f(y)2ydy+  
\int\limits_0^Vf(x)\frac{4x^2}{V}dx \cdot \int\limits_V^{\infty}f(x)2xdx) -  \\-
f(V)\int\limits_0^{\infty}f(x)\frac{(V+x)^3-|V-x|^3}{3V}xdx
\end{multline}
\subsection {Обезразмеривание уравнения Больцмана}
Для обезразмеривания уравнения (\ref{initial}) введем новые обозначения:
\newline
\begin{align}
\label{bezR}
 &V=V_0\cdot V', \,\,\, n = n_0 \cdot n', \,\,\, t = \tau_0 \cdot t' \notag\\
& V_0= (\frac{2kT_0}{m})^\frac{1}{2}, \,\,\,  f = \frac{n_0}{V_0^3} \cdot f' \notag \\
&\lambda_0 =  \frac{1}{n_0\cdot \pi d^2},\,\,\, \tau_0 = \frac{\lambda_0}{V_0}
\end{align}
где $n_0$, $T_0$ - характерные плотность и температура. Уравнение (\ref{initial}), записанное в безразмерных переменных с отброшенными штрихами, примет вид:
\begin{multline}
\label{initialF}
\frac{\partial{f(t,V)}}{\partial{t}} = 
2\pi(
 \int\limits_0^V \int\limits_{\sqrt{V^2-x^2}}^V f(x)f(y)\frac{4\sqrt{x^2+y^2-V^2}}{V}ydydx +\\+
 \int\limits_V^{\infty}f(x)2xdx \cdot \int\limits_V^{\infty}f(y)2ydy+  
\int\limits_0^Vf(x)\frac{4x^2}{V}dx \cdot \int\limits_V^{\infty}f(x)2xdx) - \\-
f(V)\int\limits_0^{\infty}f(x)\frac{(V+x)^3-|V-x|^3}{3V}xdx
\end{multline}

\subsection {Интеграл столкновений, зависящий от квадрата скорости}
Для упрощения дальнейших вычислений перейдет от интеграла столкновений, зависящего от модуля скорости, к интегралу столкновений, зависящему от квадрата модуля скорости. 
Для этого выполним замену переменных:
\begin{align}
\label{zam}
v=&V^2, \quad  p=x^2, \quad  s=y^2\notag\\
\sqrt{v}=&V, \quad \sqrt{ p}=x, \quad  \sqrt{s}=y
\end{align}

Тогда, делая подстановку из (\ref{zam}) в (\ref{initialF}) получим:
\begin{multline}
\label{initialf}
\frac{\partial{f(t,v)}}{\partial{t}} = 
2\pi (
 \int\limits_0^v \int\limits_{v-p}^v f(p)f(s)\frac{\sqrt{p+s-v}}{\sqrt{v}}dsdp + \\+
 \int\limits_v^{\infty}f(p)dp \cdot \int\limits_v^{\infty}f(s)ds +  
\int\limits_0^vf(p)\frac{\sqrt{p}}{\sqrt{v}}dp \cdot \int\limits_v^{\infty}f(s)ds + 
\int\limits_0^vf(s)\frac{\sqrt{s}}{\sqrt{v}}ds \cdot \int\limits_v^{\infty}f(p)dp)-  \\-
f(v)\int\limits_0^{\infty}f(p)\frac{(\sqrt{v}+\sqrt{p})^3-|\sqrt{v}-\sqrt{p}|^3}{6\sqrt{v}}dp
\end{multline}

\subsection {Метод конечных элементов}
Для перехода от уравнения на функцию распределения к уравнению на концентрации воспользуемся методом конечных элементов. Разобьем пространство скоростей на $N$ ячейки с центрами в точках $v_k$ с шагом $\Delta v$:
\begin{eqnarray}
\label{set}
v_k = \frac{\Delta v}{2} + k\cdot\Delta v
\end{eqnarray}
Функую распределения $f(v)$ можно представить в виде:
\begin{eqnarray}
\label{zam2}
f(v) = \sum_{k=1}^{N}f_k \cdot \delta(v-v_k)
\end{eqnarray}
 Посчитаем количество частиц $n_k$ в каждой из этих ячеек, и введем дискретную функцию распределения $f_k$  
\begin{eqnarray}
\label{zamn}
n_k = \int\limits_{v_k-\frac{\Delta v}{2}}^{v_k+\frac{\Delta v}{2}}2\pi f(v) \sqrt{v}dv = 2\pi f_k\sqrt{v_k}, \quad 
f_k = \frac{n_k}{2\pi\sqrt{v_k}}
\end{eqnarray}
Тогда, очевидно, что $f(v_k)=f_k$. В дальнейшем оба этих обозначения будут равноправными, однако для удобства мы перейдем от уравнения на функцию распределения к системе уравнений на число частиц $n_k$

\subsection {Система дискретных уравнений на $n_k$}
Для перехода от уравнения на функцию распределения к системе уравнений на плотности $n_k$, возьмем производную по времени от выражения (\ref{zamn}):
\begin{eqnarray}
\label{dN}
\frac{\partial{n_k}}{\partial{t}} = \int\limits_{v_k-\frac{\Delta v}{2}}^{v_k+\frac{\Delta v}{2}}2\pi \frac{\partial{f(v)}}{\partial{t}} \sqrt{v}dv
\end{eqnarray}
Поставляя вместо $\frac{\partial{f(v)}}{\partial{t}}$ в интеграл (\ref{dN}) правую часть уравнения (\ref{initialF}) получим систему $N$ уравнений, описывающих релаксацию газа в каждой ячейке:
\begin{multline}
\label{initialn}
\frac{\partial{n_k}}{\partial{t}} = 
 \int\limits_{v_k-\frac{\Delta v}{2}}^{v_k+\frac{\Delta v}{2}} 4\pi^2(
 \int\limits_0^v \int\limits_{v-p}^v f(p)f(s)\sqrt{p+s-v}dsdp +\\+
 \sqrt{v}\int\limits_v^{\infty}f(p)dp \cdot \int\limits_v^{\infty}f(s)ds  + 
\int\limits_0^vf(p)\sqrt{p}dp \cdot \int\limits_v^{\infty}f(s)ds + \int\limits_0^vf(s)\sqrt{s}ds \cdot \int\limits_v^{\infty}f(p)dp)- \\-
f(v)\int\limits_0^{\infty}f(p)((\sqrt{v}+\sqrt{p})^3-|\sqrt{v}-\sqrt{p}|^3)dp\,)dv
\end{multline}
где $k = 1..N$.
\newline
Главная сложность при решении этой системы уравнений состоит в подсчете тройного интеграла столкновений. В следующем разделе мы выполним его точный расчет, заменив в правой части (\ref{initialn}) функцию $f(v)$ представлением  (\ref{zam2})
 \section{Тензор столкновений}
\subsection {Расчет интеграла столкновений}
Для вычисления интеграла обратных столкновений, стоящего в правой части выражения (\ref{initialn}), разобьем его на четыре интеграла и посчитаем их отдельно. Для этого выполним замену $f(v)$ из (\ref{zam2}) и воспользуемся свойством $\delta$-функции: 
\begin{eqnarray}
\int\limits_a^b f(x)\cdot \delta(y-x) dx=\begin{cases}
f(y),a<y<b\\
0, else
\end{cases}
\end{eqnarray}
 Отдельно отметим, что вычислять пятый интеграл нет необходимости, так как его значение мы подберем исходя из законов сохранения.

\begin{multline} \nonumber
I_{1k}= 
4\pi^2 \int\limits_{v_k-\frac{\Delta v}{2}}^{v_k+\frac{\Delta v}{2}} \int\limits_0^v \int\limits_{v-p}^v f(p)f(s)\sqrt{p+s-v}\,ds\,dp\,dv = \\  = 
4\pi^2\int\limits_{v_k-\frac{\Delta v}{2}}^{v_k+\frac{\Delta v}{2}} \int\limits_0^v \int\limits_{v-p}^v
\sum_{m=1}^{N}f(v_m)\delta(p-v_m) \cdot \sum_{l=1}^{N}f(v_l)\delta(s-v_l)\sqrt{p+s-v}\,ds\,dp\,dv = \\ = 
4\pi^2\int\limits_{v_k-\frac{\Delta v}{2}}^{v_k+\frac{\Delta v}{2}}\int\limits_0^v
\sum_{m=1}^{N}f(v_m)\delta(p-v_m) \cdot \sum_{l:v_l>v-v_m}^{k}f(v_l)\sqrt{p+v_l-v}\,dp\,dv= \\ = 
4\pi^2\int\limits_{v_k-\frac{\Delta v}{2}}^{v_k+\frac{\Delta v}{2}}
\sum_{m=1}^{k}f(v_m)(\sum_{l:v_l>v-v_m}^{k}f(v_l)\sqrt{v_m+v_l-v})\,dv=  \\ =
4\pi^2\sum_{m=1}^{k}(f(v_m)\sum_{l:v_l>v_k-v_m}^{k}
(\int\limits_{v_k-\frac{\Delta v}{2}}^{v_k+\frac{\Delta v}{2}}f(v_l)\sqrt{v_m+v_l-v}\,dv))=
\end{multline} 

\begin{multline}
\label{I1}
 = 4\pi^2\frac23\sum_{m=1}^{k}(f(v_m)\sum_{l:v_l>v_k-v_m}^{k} 
(f(v_l)((v_l+v_m-v_k+\frac{\Delta v}{2})^{\frac32}-(v_l+v_m-v_k-\frac{\Delta v}{2})^{\frac32}))= \\ =
\frac23\sum_{m=1}^{k}(\frac{n_m}{\sqrt{v_m}}\sum_{l:v_l>v_k-v_m}^{k} 
(\frac{n_l}{\sqrt{v_l}}((v_l+v_m-v_k+\frac{\Delta v}{2})^{\frac32}-(v_l+v_m-v_k-\frac{\Delta v}{2})^{\frac32}))) 
\end{multline}

\begin{multline}
\label{I2}
I_{2k} =4\pi^2  \int\limits_{v_k-\frac{\Delta v}{2}}^{v_k+\frac{\Delta v}{2}} \sqrt{v}\int\limits_v^{\infty}f(p)dp \cdot \int\limits_v^{\infty}f(s)\,ds\,dv=  \\ =
4\pi^2 \int\limits_{v_k-\frac{\Delta v}{2}}^{v_k+\frac{\Delta v}{2}} \sqrt{v}
\int\limits_v^{\infty}\sum_{m=1}^{N}f(v_m)\delta (p-v_m)\, dp \cdot
\int\limits_v^{\infty}\sum_{l=1}^{N}f(v_l)\delta (s-v_l)\, ds=  \\ =
4\pi^2 \int\limits_{v_k-\frac{\Delta v}{2}}^{v_k+\frac{\Delta v}{2}}
\sum_{m=k}^N(f(v_m))\sum_{l=k}^N (f(v_l))\sqrt{v}dv=  \\ =
4\pi^2 \frac23\sum_{m=k}^N(f(v_m))\sum_{l=k}^N( f(v_l))((v_k+\frac{\Delta v}{2})^\frac32-(v_k-\frac{\Delta v}{2})^\frac32)=\\ =
4\pi^2 \frac23((v_k+\frac{\Delta v}{2})^\frac32-(v_k-\frac{\Delta v}{2})^\frac32)\cdot (\sum_{l=k}^N f(v_l))^2=\\ =
\frac23((v_k+\frac{\Delta v}{2})^\frac32-(v_k-\frac{\Delta v}{2})^\frac32)\cdot (\sum_{l=k}^N \frac{n_l}{\sqrt{v_l}})^2
\end{multline}

\begin{multline} \nonumber
I_{3k}= 4\pi^2\int\limits_{v_k-\frac{\Delta v}{2}}^{v_k+\frac{\Delta v}{2}}
( \int\limits_0^vf(p)\sqrt{p}\,dp) \cdot (\int\limits_v^{\infty}f(s)\,ds)\,dv= \\ =
4\pi^2 \int\limits_{v_k-\frac{\Delta v}{2}}^{v_k+\frac{\Delta v}{2}}
\int\limits_0^{v}\sum_{l=1}^{N}f(v_l)\delta (p-v_l)\, \sqrt{p}dp \cdot
\int\limits_v^{\infty}\sum_{m=1}^{N}f(v_m)\delta (s-v_m)\, ds=  
\end{multline}

\begin{multline}
\label{I3}
= 4\pi^2\int\limits_{v_k-\frac{\Delta v}{2}}^{v_k+\frac{\Delta v}{2}}
\sum_{l=1}^k(f(v_l)\sqrt{v_l})\cdot\sum_{m=k}^Nf(v_m)\, dv=  \\ =
4\pi^2\Delta v \cdot\sum_{l=1}^k(f(v_l)\sqrt{v_l})\cdot\sum_{m=k}^Nf(v_m)=  \\ =
\Delta v \cdot\sum_{l=1}^k(n_l)\cdot\sum_{m=k}^N (\frac{n_m}{\sqrt{v_m}})
\end{multline}

Четвертый интеграл отличается от предыдущего только индексами, поэтому повторять выкладки мы не будем и просто приведем результат. 
\begin{multline}
\label{I4}
I_{4k} = 4\pi^2\Delta v \cdot\sum_{m=1}^k(f(v_m)\sqrt{v_m})\cdot\sum_{m=l}^Nf(v_l)= \Delta v \cdot\sum_{m=1}^k(n_m)\cdot\sum_{l=k}^N (\frac{n_l}{\sqrt{v_l}})
\end{multline}
\subsection {Тензор столкновений}
Нашей целью будет записать уравнение (\ref{initialn}) в виде:
\begin{eqnarray}
\label{finaln}
\frac{\partial{n_k}}{\partial{t}} = 
 A_{ij}^kn_in_j- n_k B^k_in_i
\end{eqnarray}
Здесь по повторяющимся индексам $i и j$ ведется суммирование.
Коэффициенты $A_{ij}^k$ вычислим из интегралов, посчитанных в предыдущем разделе, а $B_i^k$ из условия сохранения числа частиц.

\subsection {Вычисление компонентов тензора столкновений}
Компоненты тензора $A_{ij}^k$ представим в виде:
\begin{eqnarray}
A_{ij}^k = A_{ij}^{1k} +A_{ij}^{2k} +A_{ij}^{3k}+A_{ij}^{4k}
\end{eqnarray}
где $A_{ij}^{1k},A_{ij}^{2k},A_{ij}^{3k},A_{ij}^{4k}$ коэффициентам пред $n_{i}n_{j}$ в интегралах  (\ref{I1}), (\ref{I2}), (\ref{I3}) и (\ref{I4}):

\begin{eqnarray}
A_{ij}^{1k}=\begin{cases}
\frac{2}{3\sqrt{v_iv_j}}((v_i+v_j-v_k+\frac{\Delta v}{2})^{\frac32}-(v_i+v_j-v_k-\frac{\Delta v}{2})^{\frac32});v_i+v_j>v_k\\
0; v_i+v_j\le v_k
\end{cases}
\end{eqnarray}

\begin{eqnarray}
A_{ij}^{2k}=\begin{cases}
\frac{2}{3\sqrt{v_iv_j}}((v_k+\frac{\Delta v}{2})^{\frac32}-(v_k-\frac{\Delta v}{2})^{\frac32});\,i\ge k \,\text{и}\, j \ge k\\
0;\,i<k \,\text{или}\, j<k
\end{cases}
\end{eqnarray}

\begin{eqnarray}
A_{ij}^{3k}=\begin{cases}
\frac{\Delta v}{\sqrt{v_j}};\,i\le k \,\text{и}\, j\ge k\\
0;\, i>k \,\text{или}\,j<k
\end{cases}
\end{eqnarray}

\begin{eqnarray}
A_{ij}^{4k}=\begin{cases}
\frac{\Delta v}{\sqrt{v_i}};\,i\ge k \,\text{и}\,j\le k\\
0;\, i<k \,\text{или}\,j>k
\end{cases}
\end{eqnarray}
Коэффициенты $A^k_{ij}$ получаются симметричными по индексам $i$ и $j$. 
\newline
Для выполнения закона сохранения числа частиц необходимо, чтобы сумма правых частей уравения по индексу k  (\ref{finaln}) давала ноль. Тогда, для коэффициентов $B^k_i $ получаем следующее выражение:
\begin{eqnarray} 
\label{B}
B^j_i = \sum^N_{k=1}A^k_{ij}
\end{eqnarray}
Для наглядной демонстрации выполнения закона сохранения частиц рассмотрим физический смысл коэффициентов $B^j_i $ и $A^k_{ij}$. 
Если между собой сталкиваются частицы из i-ой и j-ой ячейки, то число соударений за единицу времени будет равно $n_in_jB_i^j$, и все эти частицы разлетятся по k-м ячейкам. Причем, в k-ой ячейке будет $n_in_jA_{ij}^k$ частиц. А это означает, что:
$$n_in_jB^j_i = n_in_j\sum^N_{k=1}A^k_{ij}$$
Очевидно, что это равенство равносильно выражению (\ref{B}), полученному из формальных соображений.
\subsection {Корректировка тензора столкновений}
Для обеспечения выполнения закона сохранения числа частиц и энергии в системе мы должны обеспечить выполнение следующих равенств:
\begin{eqnarray}
\label{ZS0}
\begin{cases}
 \frac{\partial}{\partial{t}} \sum\limits_{k} n_k = \sum \limits_{i,j}n_i n_j \sum\limits_{k} A_{ij}^k -\sum\limits_{i,j}n_i n_j B_i^j = 0 \\
\frac{\partial}{\partial{t}} \sum\limits_{k} n_k\cdot v_k = \sum \limits_{i,j}n_i n_j \sum\limits_{k} A_{ij}^k\cdot v_k -\sum\limits_{i,j}n_i n_j B_i^j \cdot \frac{v_i+v_j}{2}= 0 
\end{cases}
\end{eqnarray}
Так как $n_i$, $n_j$, $A_{ij}^k$ и $B_i^j$ неотрицательны и условия (\ref{ZS0})должны выполняться для любых $n_i$, $n_j$, то эти условия равносильны следующим:
\begin{eqnarray}
\label{ZS1}
\begin{cases}
\sum\limits_{k} A_{ij}^k = B_i^j  \\
\sum\limits_{k} A_{ij}^k\cdot v_k =B_i^j \cdot \frac{v_i+v_j}{2} 
\end{cases}
\end{eqnarray}

Первое условие из системы (\ref{ZS1}) выполнено автоматически из определения коэффициентов $B_i^j$. А для выполнения второго условия требуется выполнить коррекцию  $A_{ij}^k$ и $B_i^j$. Скорректированные компоненты $A_{ij}^k$  будем искать в следующем виде:
\begin{eqnarray}
\label{kor}
\tilde A_{ij}^{k} = A_{ij}^{k}\cdot (1+\alpha+\beta \cdot v_k)
\end{eqnarray}
Тогда, подставляя (\ref{kor}) в (\ref{ZS1}) получим уравнения на коэффициенты $\alpha$ и $\beta$:

\begin{eqnarray}
\begin{cases}
 \sum\limits_{k} A_{ij}^k\cdot (1+\alpha+\beta \cdot v_k) = B_i^j  \\
\sum\limits_{k} A_{ij}^k\cdot (1+\alpha+\beta \cdot v_k)\cdot v_k =B_i^j \cdot \frac{v_i+v_j}{2} 
\end{cases}
\end{eqnarray}

\begin{eqnarray}
\begin{cases}
 \sum\limits_{k} A_{ij}^k+\alpha \cdot \sum\limits_{k} A_{ij}^k+\beta \cdot\sum\limits_{k} A_{ij}^k = B_i^j  \\
\sum\limits_{k} A_{ij}^k\cdot v_k+\alpha \cdot \sum\limits_{k} A_{ij}^k\cdot v_k+\beta \cdot\sum\limits_{k} A_{ij}^k\cdot v_k^2 =B_i^j \cdot \frac{v_i+v_j}{2} 
\end{cases}
\end{eqnarray}

Введем следующие обозначения:
\begin{eqnarray}
a_0 = \sum\limits_{k} A_{ij}^k,\,\,\, a_1 = \sum\limits_{k} A_{ij}^k \cdot v_k ,\,\,\, a_2 = \sum\limits_{k} A_{ij}^k \cdot v_k^2, \,\,\, b_1 = B_i^j \cdot \frac{v_i+v_j}{2}
\end{eqnarray}
Учитывая, что $B_i^j = \sum\limits_{k} A_{ij}^k$, получаем выражения для корректировочных коэффициентов:
\begin{align}
&\alpha = \frac{a_1(b_1-a_1)}{a_1^2-a_0a_2} \notag \\
&\beta = -\frac{(a_0-b_0)+\alpha a_0}{a_1}
\end{align}
Теперь, когда мы нашли скорректированный вид для тензорных компонентов $\tilde A_{ij}^k$ мы можем опустить знак $\sim$ и получить консервативную систему уравнений:
\begin{align}
\label{finalncor}
\frac{\partial{n_k}}{\partial{t}} = 
\sum\limits_{i,j} A_{ij}^kn_in_j- n_k\sum\limits_{i} B^k_in_i
\end{align}
Данная система обеспечивает выполнения законов сохранения автоматически, благодаря проведенной коррекции.
\newline
Важно подчеркнуть, что значения тензорных компонентов зависят только от разбиения и не зависят от начального распределения, поэтому подсчет коэффициентов нужно проводить только один раз, что значительно уменьшает вычислительную сложность алгоритма. В следующем разделе мы продемонстрируем работу данного метода на примере задачи о релаксации газа с различными начальными функциями распределения. 
\section{Решение задачи о релаксации}
Для демонстрации работы описанного метода будем решать задачу о релаксации газа при различных начальных условиях. 
\subsection {Общий алгоритм решения}
Построим разностную схему для решения системы уравнений (\ref{finalncor}). Проведем разбиение по времени с шагом $\Delta t$:
\begin{eqnarray}
\label{finI}
\frac{n_k^{i+1}-n_k^{i}}{\Delta t} = I_k^i
\end{eqnarray}
Верхний индекс $i$ обозначает номер временного слоя, а $I_k^i$ обозначение правой  части уравнения (\ref{finalncor}) на i-ом временном слое. 
\newline
Тогда, можем получить в явном виде выражение для плотности числа частиц на следующем временном слое:
\begin{eqnarray}
\label{nextT}
n_k^{i+1} = n_k^i+I_k^i\cdot \Delta t
\end{eqnarray}
Если нам известны значения $n_k^0$ мы сможем найти значения $n_k^i$ на любом временном слое. Количество временных слоев заранее мы определять не будем, а остановим вычисления когда значения $n_k^i$ выйдет на стационарное с необходимой точностью $\epsilon$:
\begin{eqnarray}
\label{stop}
\forall k \to |n_k^{i+1}-n_k^{i}|<\epsilon
\end{eqnarray}
\subsection {Поcтроение общего вида допустимых начальных условий}
Для проверки работоспособности предложенного метода решим задачу о релаксации газа. Для этого нужно построить начальное распределение которое сойдется к нормальному распределению:

\begin{eqnarray}
\label{max}
f(V) = n(\frac{m}{2\pi k T})^\frac{3}{2} \cdot e^{-\frac{mV^2}{2kT}}
\end{eqnarray}
В безразмерном виде (обезразмеривание аналогично пункту (\ref{bezR})) она представима в более простом виде:
\begin{eqnarray}
\label{maxBR}
f(V) = (\frac{1}{\pi})^\frac{3}{2} \cdot e^{-V^2}
\end{eqnarray}
или, в наших обозначениях:
\begin{eqnarray}
\label{maxBR}
f(v) = (\frac{1}{\pi})^\frac{3}{2} \cdot e^{-v}
\end{eqnarray}

Тогда, найдем начальное количество частиц и полную энергию системы по определению:

\begin{align}
\label{nT}
&n = 4\pi \int \limits_{0}^{\infty} f(V) \cdot V^2 dV \notag\\
&E = 4\pi \int \limits_{0}^{\infty} f(V) \cdot V^2 \cdot \frac{mV^2}{2} dV
\end{align} 

При подстановки обезразмеренной функции распределения получаем:
\begin{align}
\label{bezRnT}
&n =\frac{4}{\sqrt{\pi}}\int \limits_{0}^{\infty} e^{-V^2} \cdot V^2 dV = 1 \notag \\
&E =\frac{4}{\sqrt{\pi}}\int \limits_{0}^{\infty} e^{-V^2} \cdot V^2  \cdot V^2 dV = \frac{3}{2}
\end{align} 
Таким образом, при составлении начальных условий задачи о релаксации газа мы будем исходить из того, что $n = 1; \,\,\, E = \frac{3}{2}$.
В пространстве скоростей $v_k$ должны быть заданы дискретные начальные значения $n_k$. Зададим их в форме "ступеньки" высотой $a$, началом в ячейке с номером $b$, заканчивающуюся в ячейке с номером $c-1$. Тогда условие равенства начального числа частиц и энергии 1 и $\frac{3}{2}$ запишутся следующим образом:
\begin{align}
\label{inUsl}
&1 = \sum \limits_{k=b}^{c-1} a = a(c-b) \notag\\
&\frac{3}{2} =  \sum \limits_{k=b}^{c-1} a v_k = a\sum \limits_{k=b}^{c-1}( \frac{\Delta v}{2} + k\cdot \Delta v ) = \frac{a \Delta v (c^2-b^2)}{2}
\end{align}
От сюда получаем условия связывающие начальные параметры $a, b, c$:
\begin{align}
\label{inabc}
&c = \frac{3}{\Delta v} - b \notag \\
&a = \frac{1}{\frac{3}{\Delta v} - 2b}
\end{align}
В следующем разделе мы приведем примеры расчета задачи о релаксации газа при различных параметрах "ступеньки"

\subsection {Примеры расчета задачи о релаксации при различных начальных условиях}
Разобьем пространство скоростей на $N = 100$ ячеек шириной $\Delta v = 0,05$ и будем проводить релаксацию с шагом по времени $\Delta t = 0,01$. Выход на стационарное состояние с относительной погрешностью $<10^{-2}$ достигается при $t = 3$.
Для демонстрации работы метода приведем графики показывающие релаксацию газа при различных начальных условиях. Для этого выбираем различные параметры $b$ в ступеньке  и рассчитываем допустимую "ширину" и "высоту". 
\newline
\vspace{2em}
\includegraphics[width=12cm]{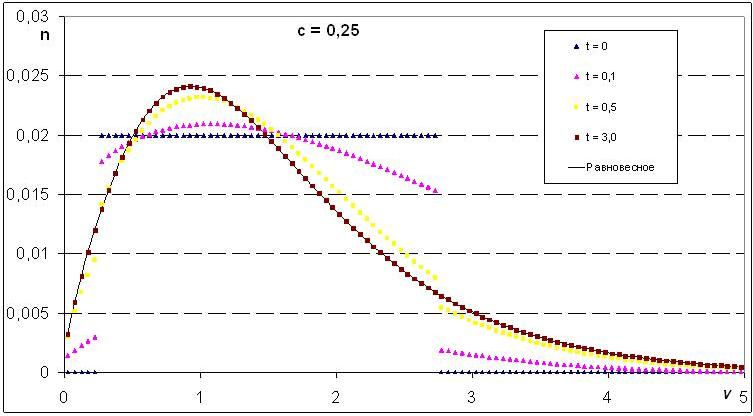} 
\newline
\vspace{2em}
\includegraphics[width=12cm]{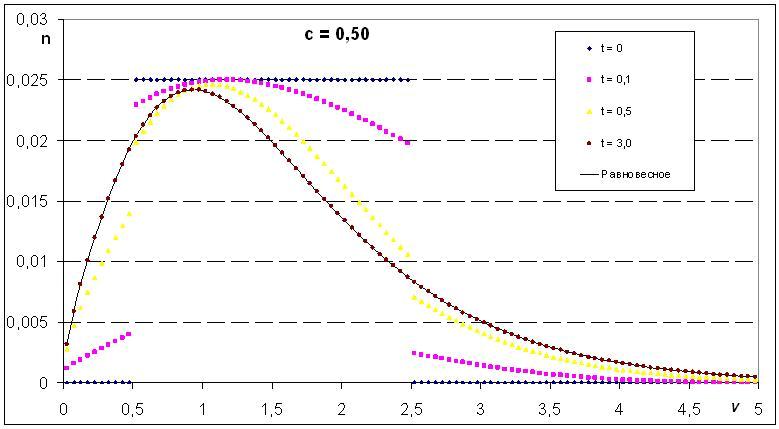} 
\newline
\vspace{2em}
\includegraphics[width=12cm]{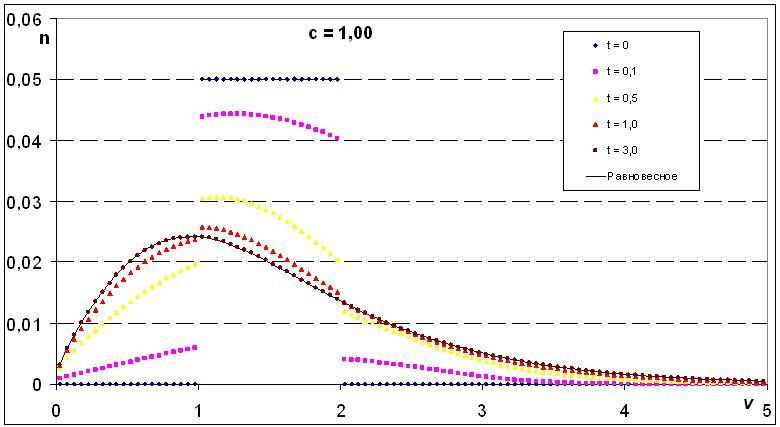} 
\newline

В заключение, приведем график равновесной функции рапределения $f_k = \frac{n_k}{2\pi\sqrt{v_k}}$ в стандартных координатах $V = \sqrt{v}$:
\newline
\includegraphics[width=12cm]{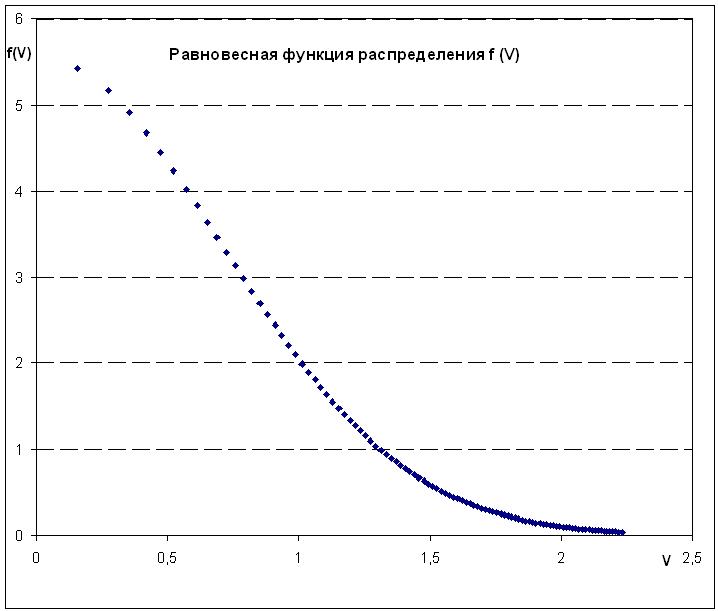} 
\newline
Как видно из приведенных расчетов, предложенный метод работоспособен так как выходит на стационарное распределение, совпадающее с нормальный распределенным.
\section{Заключение}
В работе был предложен метод решения кинетического уравнения Больцмана, основанный на построении его дискретной консервативной модели. Дискретный аналог интеграла столкновений был представлен в виде свертки тензора столкновений, независящего от функции начального распределения, с тензором с компонентами из средних плотностей в ячейках. Построенный таким образом дискретный оператор столкновений обладает свойством консервативности. Численная реализация дискретной модели на задаче об изотропной релаксации газа показывает, что метод обладает высоким быстродействием. Численный метод реализован для модели твердых сфер.
\newline
Главной особенностью данного метода является независимость компонент тензора столкновений от функции распределения. Таким образом, компоненты тензора столкновений вычисляются один раз для разных начальных функций распределения, что значительно увеличивает быстродействие предложенного метода.
\newline
Предложенный метод может быть использован при решении куда более сложных задач: задача релаксации смеси газов, химически реагирующих смесей газом или системы с возбуждаемыми уровнями энергий частиц. Таким образом, данная работа имеет фундаментальное значение для решения задач кинетической теории газов.


\newpage
\begin{thebibliography}{99}
\bibitem {baran} Баранцев Р.Г. Об ударных трасформантах кинетического уравнения аэродинамики разреженных гахов // Сб. Аэродинамика разреженных гахов. Л.: Издательство ЛГУ. 1963. Вып. 1. с. 80-91
\bibitem {aristov} Аристов В.В. О решении уравнения больцмана для дискретных скоростей // ДАН СССР. 1985. т. 283. №4. с.831-834
\bibitem {rykov} Рыков В.А. Релаксация газа, описымаевого кинетическим уравнением Больцмана // Прикладная математика и ханика 1967. т. 31. вып. 4. с. 756-762
\end{thebibliography}
 \end{document}